\documentclass[aps,prb,amsmath,amssymb,twocolumn,superscriptaddress,floatfix,longbibliography]{revtex4-2}

\usepackage{graphicx}
\usepackage{subfigure}
\usepackage{booktabs}
\usepackage{float}
\usepackage[pdftex,colorlinks=true,hypertexnames=false]{hyperref}
\hypersetup{pdfpagemode=UseNone,pdfpagelayout=OneColumn,pdfstartview=FitH}

\begin{document}

\title[]{Magnetic anisotropy and dipolar interactions in the frustrated triangular-lattice magnet NaGdS$_2$}

\author{J. Grumbach}
\email{E-Mail: Justus.Grumbach@tu-dresden.de}
\affiliation{Institut f\"ur Festk\"orper- und Materialphysik, Technische Universit\"at Dresden, 01062 Dresden, Germany}

\author{E. H\"außler}
\affiliation{Fakult\"at f\"ur Chemie und Lebensmittelchemie, Technische Universit\"at Dresden, 01062 Dresden, Germany}

\author{S. Luther}
\affiliation{Hochfeld-Magnetlabor Dresden (HLD-EMFL) and W\"urzburg-Dresden Cluster of Excellence ct.qmat, Helmholtz-Zentrum Dresden-Rossendorf, 01328 Dresden, Germany}

\author{J. Sichelschmidt}
\affiliation{Max-Planck-Institut f\"ur Chemische Physik fester Stoffe, 01187 Dresden, Germany}

\author{K. M. Ranjith}
\affiliation{Max-Planck-Institut f\"ur Chemische Physik fester Stoffe, 01187 Dresden, Germany}

\author{T. Herrmannsd\"orfer}
\affiliation{Hochfeld-Magnetlabor Dresden (HLD-EMFL) and W\"urzburg-Dresden Cluster of Excellence ct.qmat, Helmholtz-Zentrum Dresden-Rossendorf, 01328 Dresden, Germany}

\author{M. Rotter}
\affiliation{$McPhase~project$, 01159 Dresden, Germany}

\author{S. Granovsky}
\affiliation{Institut f\"ur Festk\"orper- und Materialphysik, Technische Universit\"at Dresden, 01062 Dresden, Germany}

\author{H. K\"uhne}
\affiliation{Hochfeld-Magnetlabor Dresden (HLD-EMFL) and W\"urzburg-Dresden Cluster of Excellence ct.qmat, Helmholtz-Zentrum Dresden-Rossendorf, 01328 Dresden, Germany}

\author{M. Uhlarz}
\affiliation{Hochfeld-Magnetlabor Dresden (HLD-EMFL) and W\"urzburg-Dresden Cluster of Excellence ct.qmat, Helmholtz-Zentrum Dresden-Rossendorf, 01328 Dresden, Germany}

\author{J. Wosnitza}
\affiliation{Institut f\"ur Festk\"orper- und Materialphysik, Technische Universit\"at Dresden, 01062 Dresden, Germany}
\affiliation{Hochfeld-Magnetlabor Dresden (HLD-EMFL) and W\"urzburg-Dresden Cluster of Excellence ct.qmat, Helmholtz-Zentrum Dresden-Rossendorf, 01328 Dresden, Germany}

\author{H.-H. Klauß}
\affiliation{Institut f\"ur Festk\"orper- und Materialphysik, Technische Universit\"at Dresden, 01062 Dresden, Germany}

\author{M. Baenitz}
\affiliation{Max-Planck-Institut f\"ur Chemische Physik fester Stoffe, 01187 Dresden, Germany}

\author{T. Doert}
\affiliation{Fakult\"at f\"ur Chemie und Lebensmittelchemie, Technische Universit\"at Dresden, 01062 Dresden, Germany}

\author{M. Doerr}
\affiliation{Institut f\"ur Festk\"orper- und Materialphysik, Technische Universit\"at Dresden, 01062 Dresden, Germany}

\begin{abstract}
In this comprehensive study, we present results of bulk measurements (magnetization, specific heat, ac susceptibility, thermal expansion, and magnetostriction) combined with local methods such as nuclear magnetic resonance ($^{23}$Na NMR) and electron spin resonance (ESR) and simulations (\textit{McPhase}) on polycrystalline and single-crystalline NaGdS$_2$ samples. The rare-earth delafossite NaGdS$_2$ is a triangular-lattice magnet with $S$ = 7/2  spin-only Gd$^{3+}$ moments with suppressed single-ion anisotropy. 
In our study, we estimate that NaGdS$_2$ has a weak antiferromagnetic exchange ($J_H/k_B \approx$ 52\,mK) and signs of long-range magnetic order are absent down to lowest temperature. However, indications of short range magnetic order are found below 180\,mK in the ac susceptibility and thermal expansion. Our results indicate an interplay of Heisenberg-type and dipolar exchange. Due to the large moment of the Gd$^{3+}$ ions, one expects a strong impact of the dipolar coupling in NaGdS$_2$, in contrast to the related NaYbS$_2$. ESR and $^{23}$Na NMR measurements, indeed, indicate the formation of short-range ferromagnetic correlations. NaGdS$_2$ appears to be a rare system, in which magnetic order is suppressed by a competition between Heisenberg and dipolar interactions.
\end{abstract}

\maketitle


\section{Introduction}
Selected, mostly planar, lattices with a defined spin arrangement (honeycomb, triangle, square) are suitable to realize a spin-liquid (SL) state. In addition to the crystal structure, the type of exchange interaction plays a decisive role. In recent years, some 4$f$-based quantum magnets have emerged as excellent candidates for the research of spin-liquids and other unusual magnetic states, as some Kramers ions host a quasi spin-1/2 state at low temperatures. 

\begin{figure}[t]
    \centering
    \includegraphics[width=0.48\textwidth]{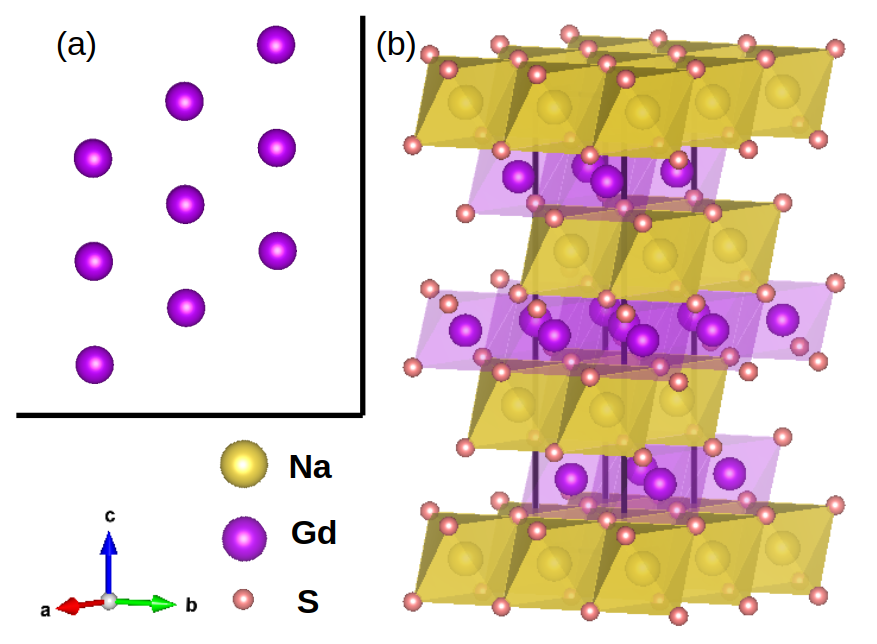}
    \caption{Crystal structure of NaGdS$_2$ with (a) hexagonal layers of Gd$^{3+}$ ions viewed along [001] and (b) alternating stacking of nonmagnetic and magnetic planes of NaS$_6$ and GdS$_6$ octahedra. The visualization was done in \textit{VESTA} \cite{vesta3}. }
    \label{fig:vesta}
\end{figure}

Promising candidates for quantum spin-liquid materials based on triangular lattices are rare-earth delafossite compounds ($\alpha$-NaFeO$_2$ structure type) of composition $ARX_2$. Thereby $A$ denotes a monovalent ion (mostly an alkali metal), $R$ denotes a trivalent rare-earth ion, and $X$ stands for a chalcogenide. Delafossites crystallize in the centrosymmetric trigonal \textit{R}$\overline{3}$\textit{m} space group. 
The structure is built out of a two-dimensional (2D) network of $AX_6$ and $RX_6$ octahedra [see Fig. \ref{fig:vesta}(b)] \cite{Marquardt05, Miyasaka09}. Ideal triangular rare-earth planes are separated by nonmagnetic layers [see Fig. \ref{fig:vesta}(a)], providing delafossites as triangular-lattice quantum magnets. 

Most prominent members of the compound family are NaYb$X_2$, where the spin-liquid ground state seems to be well achieved \cite{BSchmidt_2021}.
In other cases, several types of antiferromagnetically ordered compounds were found. The list of compounds is constantly growing. For instance, NaYbO$_2$ \cite{Ranjith_nyo} NaYbS$_2$ \cite{ysbaenitz, yssarkar}, NaYbSe$_2$ \cite{Ranjith_yse, yse}, KYbS$_2$ \cite{kys},  CsCeS$_2$, CsYbSe$_2$ \cite{cs}, and KHoSe$_2$ \cite{Boswell_KHSe} display possible quantum spin liquid ground states, whereas KCeS$_2$ \cite{cesbastien, ceskulbakov} and KCeSe$_2$ \cite{cese} show long-range magnetic order. In addition, the non-Kramers-ion material KTmSe$_2$ \cite{Zheng_KTS} was also investigated, also revealing absent long range order. 

Here, we report on the rare-earth delafossite NaGdS$_2$. This compound is a triangular-lattice magnet with 4$f^{7}$ $S$ = 7/2 spin-only Gd$^{3+}$ moments without single-ion anisotropy. Indeed, in Gd compounds often weaker influences such as dipolar contributions or effects due to the crystal symmetry affect anisotropies. So far, no magnetic order has been reported for NaGdS$_2$, which means that it is a possible spin liquid. This also favors the system as a superior cooling material for adiabatic demagnetization applications \cite{ngdelacotte}. However, due to the large spin moment and less spin-orbit entanglement, one would actually expect order. In contrast, other possible Gd-based spin liquids have a more three-dimensional exchange network \cite{paddison, petrenko}. 
In the 4$f$-ion relative NaYbS$_2$ the absence of magnetic order and the spin-liquid ground state originate from bond-dependent frustration due to spin-orbit entanglement \cite{BSchmidt_2021}. 
%

In this context, the question arises, which interactions compete
with each other in NaGdS$_2$ and what additional influence the geometric frustration
of the triangular lattice has. For temperatures above 1\,K, the magnetic
exchange (Heisenberg) is the decisive interaction, whereas in the mK
temperature range the dipolar interaction can also become important.
The latter is often much weaker than the magnetic exchange and is only used as a correction variable that extends the range of the coupling strength of the nearest-neighbor interactions $J_1$ and $J_2$, which leads to spin liquid behavior \cite{Keles_2018}. Of particular interest are, therefore, compounds in which the dipole exchange is in the similar energy range as the exchange coupling. 
Especially for Gd$^{3+}$ compounds with large spin moment stronger dipolar effects are expected compared to Yb$^{3+}$ compounds, since the interactions are quadratic with the magnetic moment. Theoretical work on compounds with triangular lattice also points to the importance of dipolar ordering especially for Gd$^{3+}$ compounds \cite{Keles_2018, Yao_Dipolar}. Also experimental work on triangular lattices with Gd$^{3+}$ discusses the role of dipolar interactions. One example is KBaGd(BO$_3$)$_2$, where magnetic order was detected at $T_{\rm N}$ = 262\,mK \cite{JescheGd22, xiang2023dipolar}. This compound family seems to be a good example for very weak exchange interaction; even a related Yb$^{3+}$ compound orders via dipolar interactions \cite{BagBYB}. Our compound NaGdS$_{2}$ has the advantage that 
randomness of the ion distribution can be neglected, so
that the presence of distributed exchange couplings should be far less significant.

\section{Experimental}
\subsection{Synthesis}

We grew NaGdS$_2$ powder samples and crystals at 1050\,°C from a Na$_2$CO$_3$ or Na$_2$S flux, respectively, in a protective argon atmosphere by a slightly modified procedure as reported by Masuda \emph{et al.} \cite{masuda_1999}.

For the powder samples, we mixed dried Na$_2$CO$_3$ (4239.6\,mg, 40\,mmol, 40\,eq.) with 1 eq. of Gd$_2$O$_3$ (Alfa Aesar, $99.999$\,\%) in a porcelain mortar and filled in a glassy carbon boat, which was placed in a ceramic tube inside a tubular furnace. 
After flushing all pipes and a 1\,l flask containing liquid CS$_2$ with argon for 30 min, we heated the furnace up to 1050\,°C within three hours ($\approx 5.7$ K/min) under a stream of argon of approx. $2$\,l/h. While dwelling for 30 minutes, CS$_2$ was transported to the reaction zone by bubbling argon ($\approx$ 5\,l/h) through the reservoir.
Under an unloaded stream of argon, we cooled the furnace to 600\,°C within three hours (2.5\,K/min) and without further control to room temperature.
After dissolving the solidified flux with distilled water, we filtered off the target compound and washed it with ethanol. 
Leftovers of small black particles, presumably carbon, were decanted after ultrasonic treatment in ethanol. 
The remaining product was found to consist of mostly intergrown agglomerates of small colorless hexagonal crystals (1-10 \,$\mu$m).

\begin{figure}[t]
    \centering
    \includegraphics[width=0.48\textwidth]{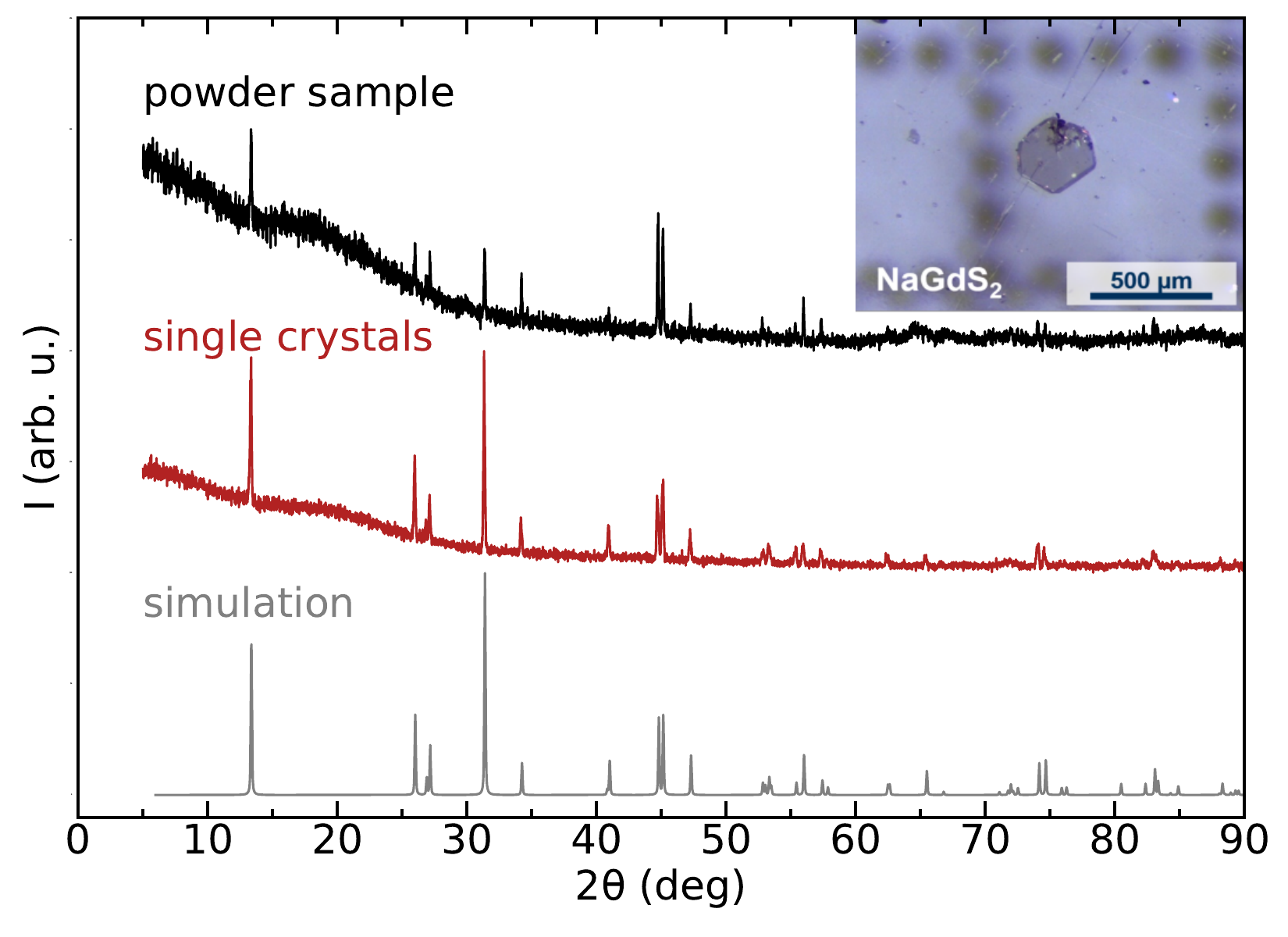}
    \caption{Powder diffraction patterns of the two NaGdS$_2$ batches used in this study. Selected single crystals were ground to powder to measure the spectrum. One of the transparent single crystals is shown in the inset.}
    \label{fig:NaGdS2 crystal}
\end{figure} 

For growing larger crystals up to $800$\,$\mu$m in diameter, we used dried Na$_2$S as flux with the same molar ratio in the mixture.
During the synthesis, the dwelling time was prolonged to one hour, and the cooling rate was decreased to 1.25\,K/min. One of the resulting crystals is shown in the inset of Fig. \ref{fig:NaGdS2 crystal}.
We evaluated the phase purity of the samples by x-ray powder diffraction measured on a Stoe Stadi P diffractometer using Cu-K$\alpha$$_1$ radiation, also shown in Fig. \ref{fig:NaGdS2 crystal} 

\subsection{Measurement methods}
We performed magnetization measurements on the powder samples using a SQUID magnetometer down to 0.5\,K providing a general magnetic characterization. 
For our measurements on single crystals, we used a VSM system in a cryo-free system (CFMS), which operates as a liquid-helium closed-cycle system, between 1.5 and 300\,K and magnetic fields up to 12\,T. We applied sweep rates of 0.1\,T/min to 0.3\,T/min and 0.1\,K/min to 1\,K/min to assure full thermalization of the samples. The plate-like single crystals were mounted with the basal plane on two different non-magnetic holders in order to do measurements in-plane and out-of-plane.

We measured the magnetic susceptibility down to 60\,mK using an ac-method by employing a compensated pick-up coil system installed in a $^3$He/$^4$He dilution refrigerator, capable of reaching fields up to 16\,T. The sweep rate of the temperature was between 1 and 1.5\,mK/min.

For temperature- and field-dependent dilatometry measurements, we used a capacitive tilted-annulus dilatometer cell with a sensitivity to relative length changes of $\sim$10$^{-7}$ \cite{Dilatometer} in a dilution refrigerator.

We measured the specific heat in the mentioned CFMS using an ac method. A single crystal of a mass of 40\,$\mu$g was glued to a 100x100\,$\mu$m$^2$ membrane, which is equipped with a heater. For measuring the temperature, we used a semiconducting \textit{Cernox} resistor.

Further, we investigated the electron spin resonance (ESR) of a NaGdS$_{2}$ single crystal.
We probed the ESR by detecting the power $P$ absorbed by a transversally applied magnetic microwave field.
For improved signal-to-noise ratio, we recorded the first derivative of the absorbed microwave power, $\mathrm{d}P/\mathrm{d}H$. We used a continuous-wave ESR spectrometer at the X-band (9.4 GHz) and employed a He-flow cryostat to reach temperatures down to 4\,K. The measured resonance spectra were described with a Lorentzian shape including the influence of the counter-rotating component (``counter resonance") of the linearly polarized microwave field \cite{ehlers19a,rauch15a}. From this fit, we obtained the linewidth $\Delta H$ and resonance field $H_{\rm res}$. 

We used a conventional pulsed NMR method on the $^{23}$Na nuclei (nuclear spin $I=3/2$)
in the temperature range between 2 and 295\,K, at a fixed frequency of 35\,MHz. We mixed NaGdS$_2$ powder with paraffin, heated it up, shook it, and then cured it to randomize the grains.
The field-sweep spectra are obtained by integration over the spin
echo in the time domain. The powder spectra are
modelled with an anisotropic shift tensor. We determined tensor components in the $ab$ plane and along the $c$ direction.

\section{Results}
\subsection{Magnetization}
\begin{figure}[H]
    \includegraphics[width=\linewidth]{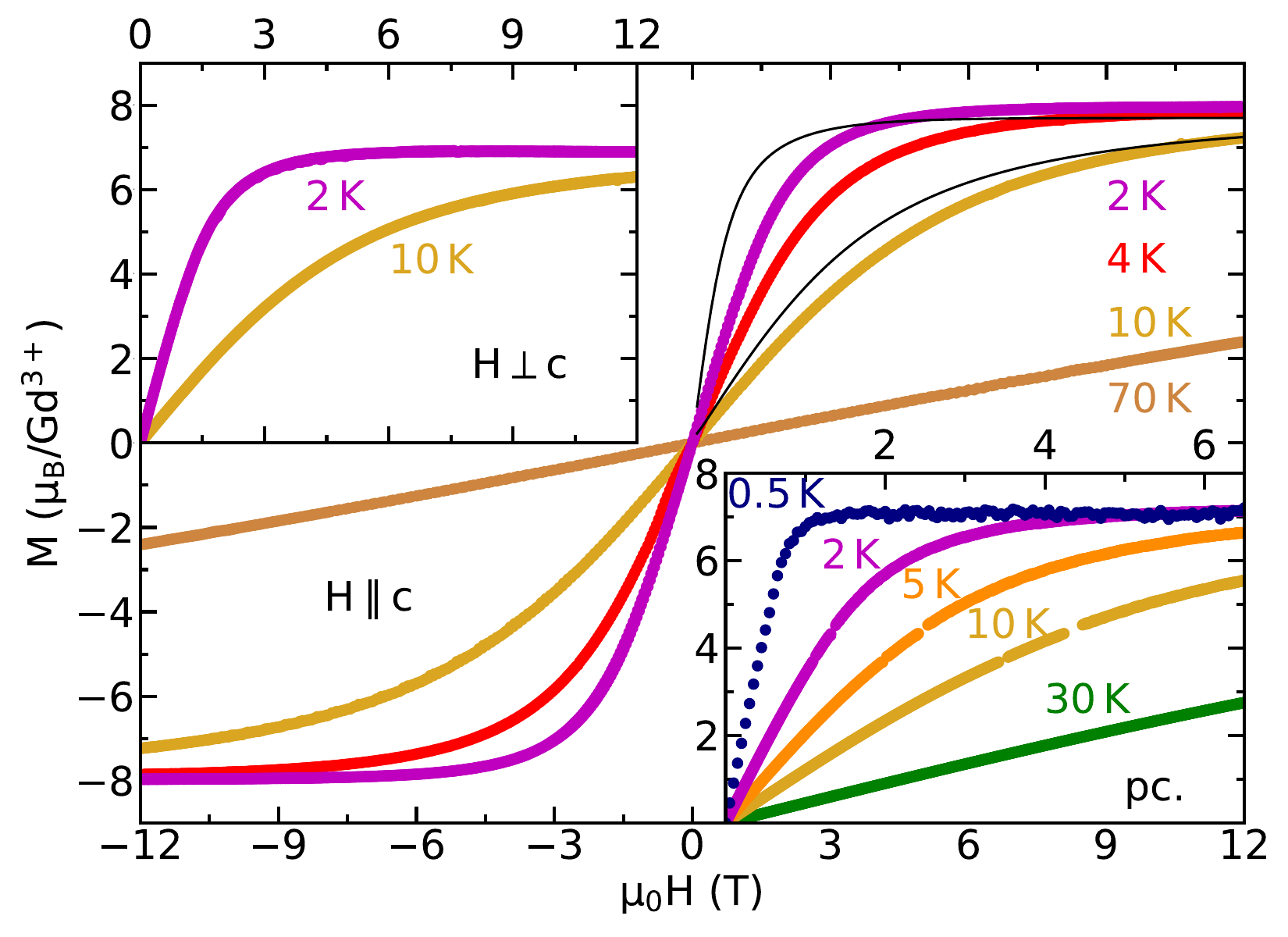}
    \caption{Field-dependent magnetization of single-crystalline NaGdS$_2$ for $H\parallel c$ (main figure) and $H\perp c$ (left inset) as well as polycrystalline NaGdS$_2$ (right inset) at various temperatures. The solid lines show Brillouin functions. }
    \label{mag}
\end{figure}

\begin{figure*}[t]
    \centering
    \setkeys{Gin}{width=0.48\textwidth}
    \subfigure{\includegraphics[width=0.48\textwidth]{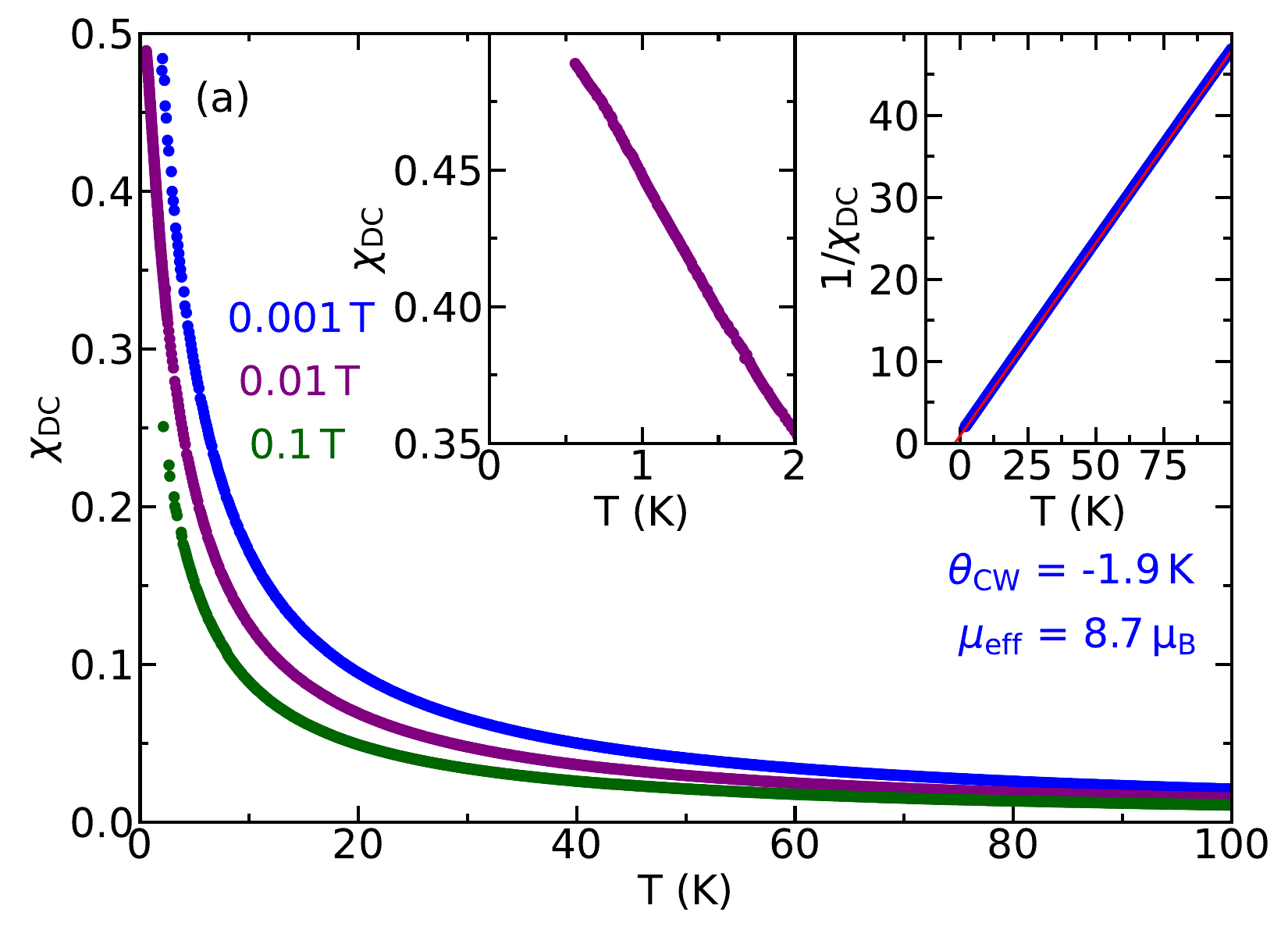}}
    \subfigure{\includegraphics[width=0.48\textwidth]{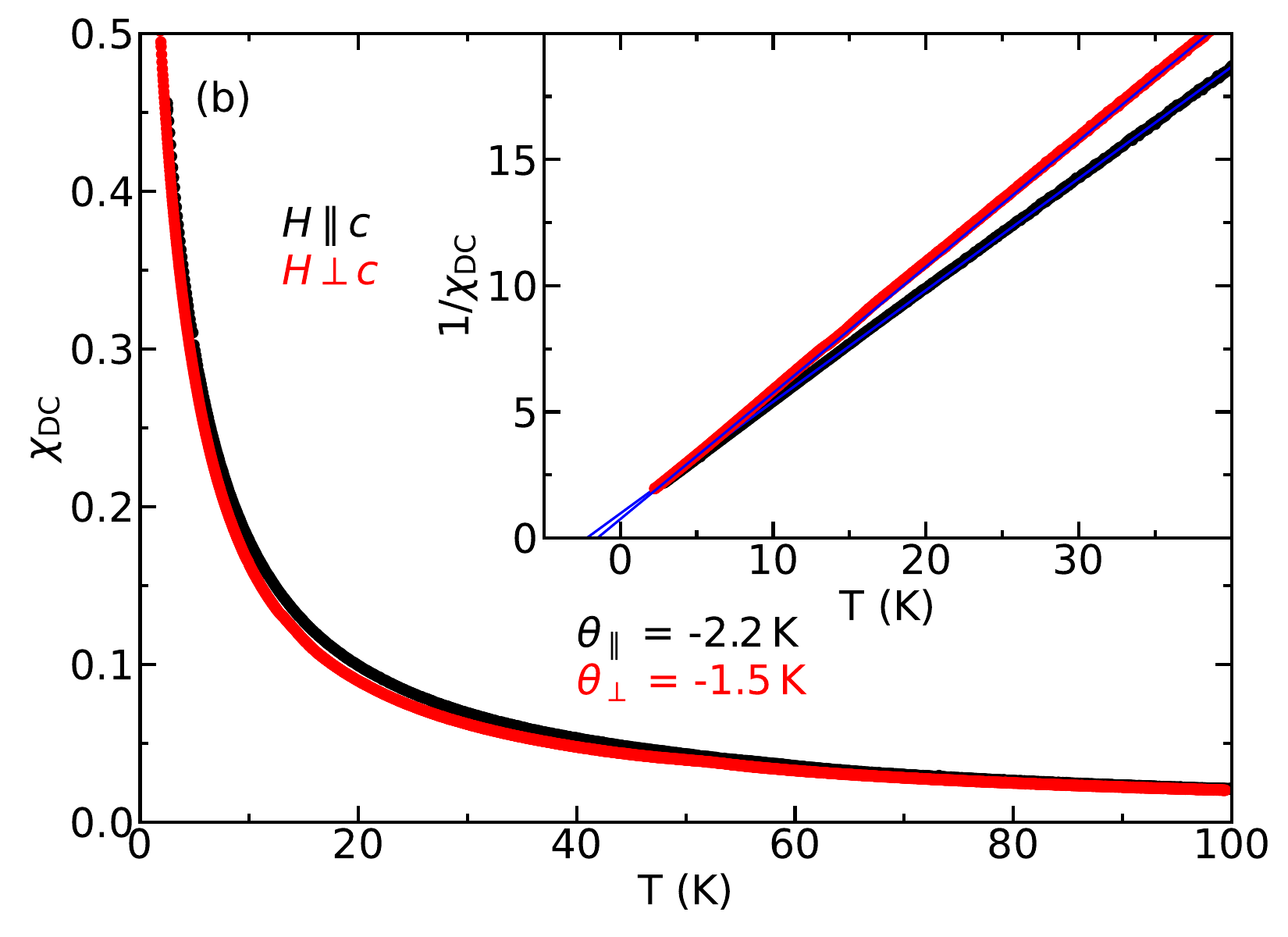}}
    \caption{Temperature dependence of the magnetic dc susceptibility of (a) polycrystalline material and (b) single crystals of NaGdS$_2$. The left inset in (a) shows an enlargement of the data in $\mu_0H$ = 0.01\,T below 2\,K revealing the absence of magnetic order down to 0.5\,K.
    The right inset in (a) and the inset in (b) show the inverse susceptibility and Curie-Weiss fits (solid lines) indicating paramagnetism and revealing a small anisotropy in the single crystal (b). }
    \label{sus}
\end{figure*}

The magnetization of the polycrystalline sample [see right inset of Fig. \ref{mag} and Fig. \ref{sus}(a)] shows the absence of magnetic order down to 0.5\,K, a saturation moment of 7.2\,$\mu_{\rm B}$ at a saturation field of 1\,T at 0.5\,K (in contrast to about 15 T\,for NaYbS$_2$ \cite{ysbaenitz}) and a rather small negative Curie-Weiss temperature of -1.9\,K [see Curie-Weiss fit to the dc-susceptibility data in the right inset of Fig. 4(a)]. The negative Curie-Weiss temperature indicates an overall dominant antiferromagnetic exchange between the moments, consistent with predictions \cite{plug}.

We further investigated the magnetic anisotropy by measuring the magnetizations for fields aligned in plane ($H\perp c$) and out of plane ($H\parallel c$), shown in Fig. \ref{mag}. The saturation magnetization for $H\parallel c$ (7.96\,$\rm \mu_B$) is slightly larger, than for $H\perp c$ (6.90\,$\rm \mu_B$). From Curie-Weiss fits to the dc-susceptibility data, we determined very similar, but larger Curie-Weiss moments of $\mu_{\rm eff}^{\parallel}$ = 8.6\,$\mu_{\rm B}$ and $\mu_{\rm eff}^{\perp}$ = 8.8\,$\mu_{\rm B}$. This, in company with the saturation magnetizations may be caused by modified $g$-factors in the actual crystal, which motivated our ESR measurements.
For reference, the effective Curie-Weiss moment for the free Gd$^{3+}$ ion is given by
$\mu_{\rm eff} = g\sqrt{J(J+1)}\,\mu_{\rm B} = 7.94\,\mu_{\rm B}$ (assuming $g \approx$ 2), whereas the saturation moment in the magnetization is $M_{\rm sat} = g_{\rm J}J\mu_{\rm B} = 7.01\,\mu_{\rm B}$.
 
Interestingly, also the Curie-Weiss temperatures differ slightly for the two directions ($\theta_{\parallel}$ = -2.2\,K and $\theta_{\perp}$ = -1.5\,K) [see Fig. \ref{sus}(b)]. A small anisotropy exceeds the respective error bars of $\pm$0.2\,K. Assuming only Heisenberg interactions, this indicates a larger coupling of the moments along the $c$ direction than in the $ab$ plane. This is in contrast to NaYbS$_2$, for which a strong anisotropy exists with $\theta_{\parallel}$ = -4.5\,K and $\theta_{\perp}$ = -13.5\,K \cite{ysbaenitz}. This points towards a different origin of the anisotropy. While for NaYbS$_2$ crystal-field effects dominate, for NaGdS$_2$ dipolar interactions are of major importance.
The stronger dipolar interaction in the plane, due to shorter interatomic distances, might be the reason for the stronger reduction of the Curie-Weiss temperature when the magnetic field is oriented in-plane.

We further made a comparison of our magnetization data to paramagnetic Brillouin functions for $H\parallel c$ (Fig. \ref{mag}). For that we used the $g$-factor obtained by ESR, as discussed later. The differences are only small, with delayed polarizations (similar as reported in \cite{JescheGd22}), which hint at the presence of only weak antiferromagnetic exchange contributions. Hence, down to 2\,K, we can describe NaGdS$_2$ as a paramagnet with weak exchange and small anisotropy.

\subsection{Specific heat}
\begin{figure}[b]
  \begin{center}
  \includegraphics[width=\linewidth]{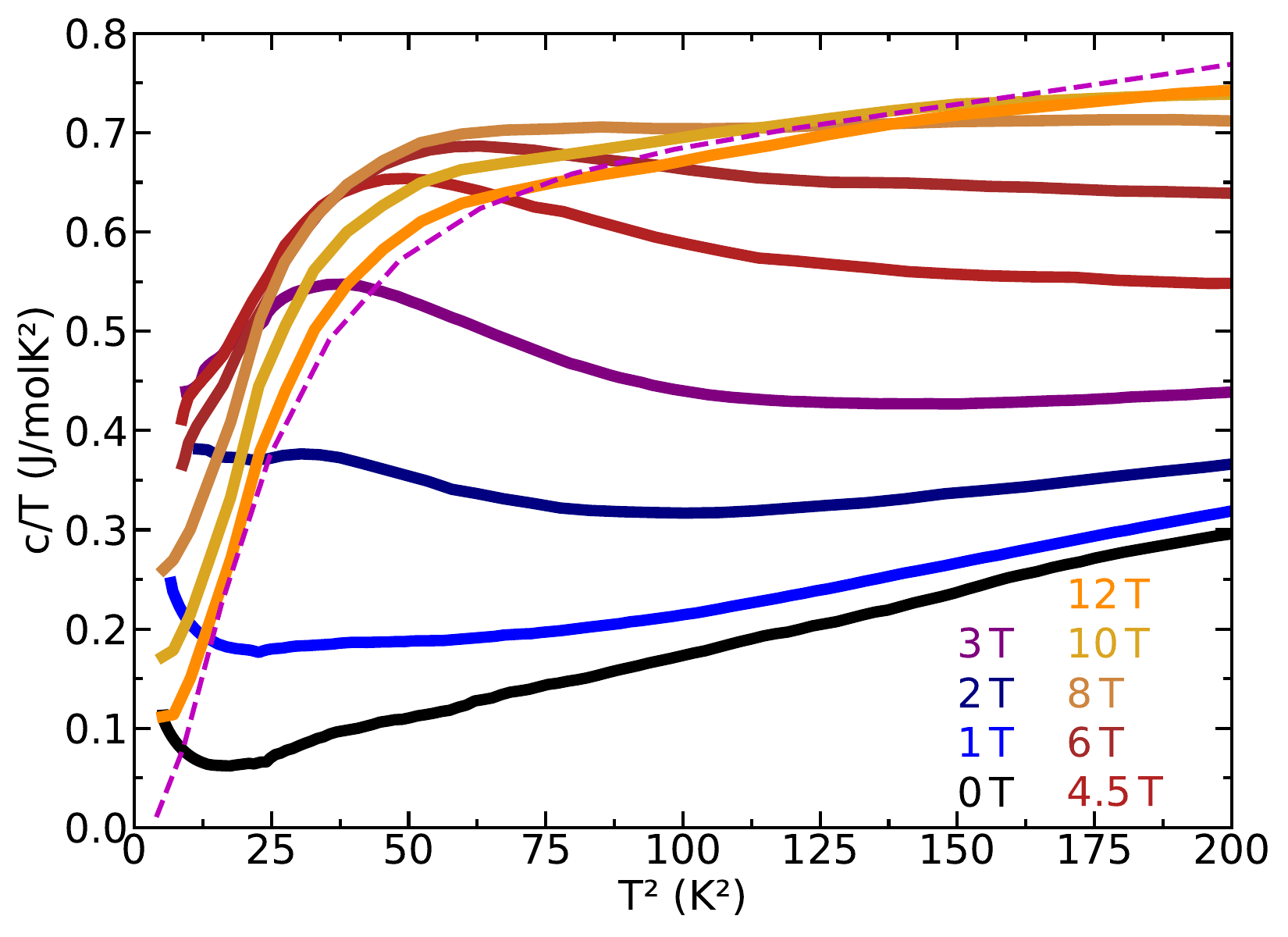}    
  \end{center}
  \caption{Specific heat divided by $T$ vs. $T^2$ at different magnetic fields. In zero and low magnetic fields the upturn at lowest temperatures indicates magnetic contributions. At higher fields a broad Schottky-like anomaly develops. The dashed line shows the specific heat calculated for a 8-level Schottky anomaly at 12 T.}
  \label{Cap}
\end{figure}

We measured the specific heat as function of temperature at various magnetic fields up to 12\,T. The data at zero field follow a classical Debye-like behavior and reach the Dulong-Petit value at about 200\,K (data not shown). No sharp feature hinting at a phase transition is present above 2\,K. In zero field, we observe an upturn in c/T below 4\,K (Fig. \ref{Cap}). This indicates an onset of magnetic correlations. In applied magnetic fields, this onset is gradually disappearing and a broad magnetic contribution appears at 1\,T and higher fields, shifting to higher temperatures with increasing fields. This anomaly can be explained assuming 8-level Schottky anomalies of the Gd$^{3+}$ S = 7/2 ground state. 

For multi-level systems, an equation of the energy of n degenerate levels can be derived to be 
\begin{equation}
    E = N\,\frac{\sum\limits_{r=0}^{n-1} E_r\,exp\left(-\frac{E_r}{k_B\,T}\right)}{\sum\limits_{r=0}^{n-1} exp\left(-\frac{E_r}{k_B\,T}\right)},
\end{equation}
from which the contribution to the specific heat results by differentiation with respect to the temperature.

In our case we assumed equidistant energies, Zeeman-split by the external magnetic field. We neglected inner magnetic fields, assuming this effect to be low as deduced before. The resulting Schottky-curve is examplary shown for the data at 12\,T (Fig. \ref{Cap}). Despite the simplicity of the model, the peak is resembled reasonably well, showing that the Schottky-anomaly due to Zeeman-splitting is the main contribution of this peak.

\subsection{Electron spin resonance \label{ESR}}

\begin{figure}[H]
\begin{center}
\includegraphics[width=0.8\columnwidth]{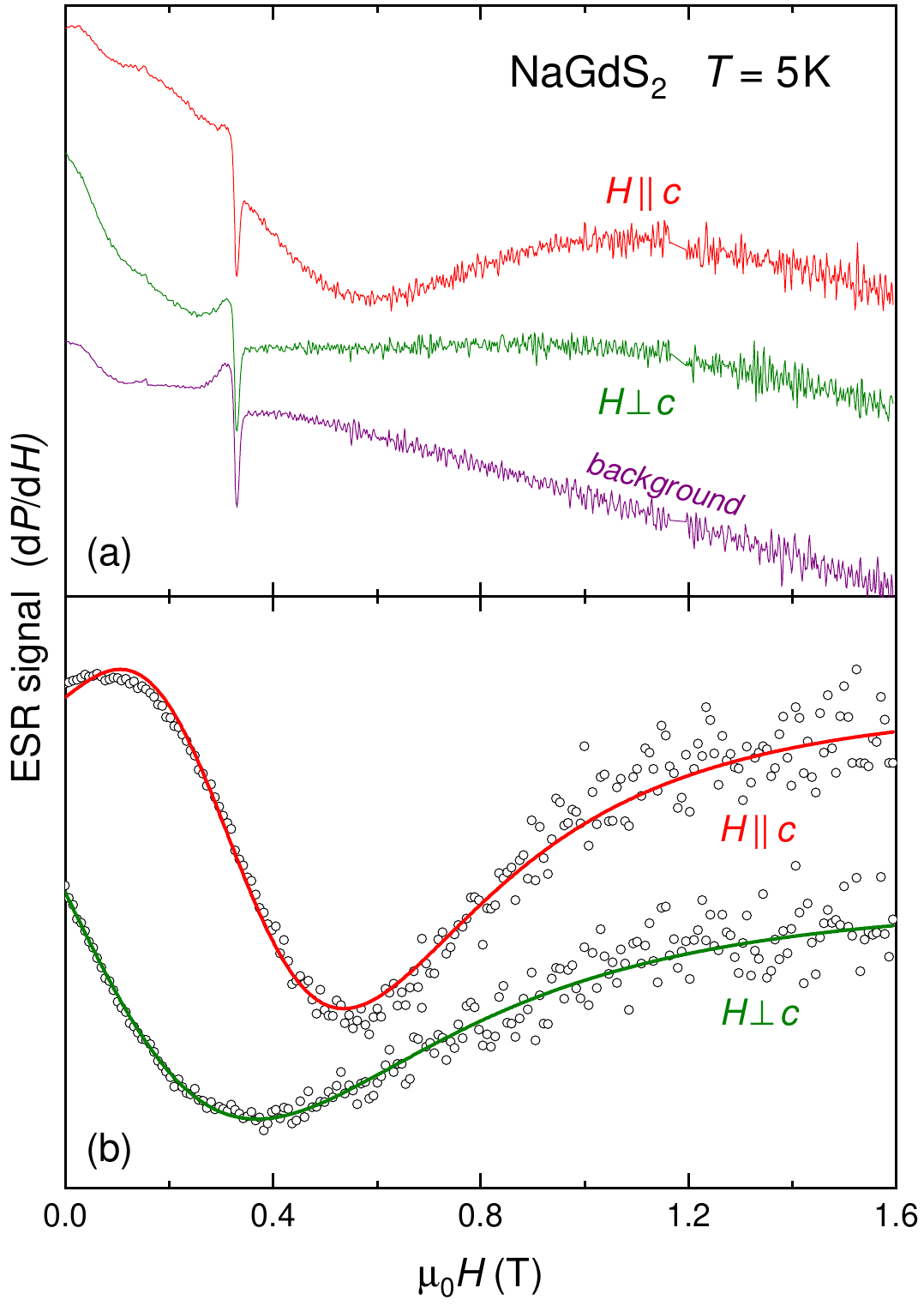}
\end{center}
\caption{Typical ESR spectra of NaGdS$_{2}$ at 5\,K. (a) Raw ESR spectra for in-plane ($H\bot c$) and out-of-plane ($H\|c$) external field directions including a prominent background.
(b) ESR spectra after subtracting the background. Solid lines denote Lorentzian line shapes which also take the counter resonance into account.
}
\label{FigESR1}
\end{figure}

We investigated the ESR of a small ($m\simeq0.02\,\rm mg$) NaGdS$_{2}$ single crystal. 
As shown in Fig. \ref{FigESR1}(a) the obtained spectra are weak as compared to the background, and broad in field. We found such spectra for temperatures up to about 60\,K, above which the sample signal reaches the detection limit. Prior fitting with a symmetric Lorentzian line shape (solid lines), we subtracted the background yielding the spectra as displayed in Fig. \ref{FigESR1}(b). This finding for the shape is in contrast to the non-Lorentzian shape reported in a previous ESR study of Gd$^{3+}$ spins on a triangular lattice for TlGdSe$_{2}$ \cite{duczmal03a}.
We determined the ESR $g$ values via the resonance field $H_{\rm res}$ which appears to be strongly anisotropic. 
This points to internal fields adding to the external field most effectively in the plane. 

\begin{figure}[H]
\begin{center}
\includegraphics[width=0.877\columnwidth]{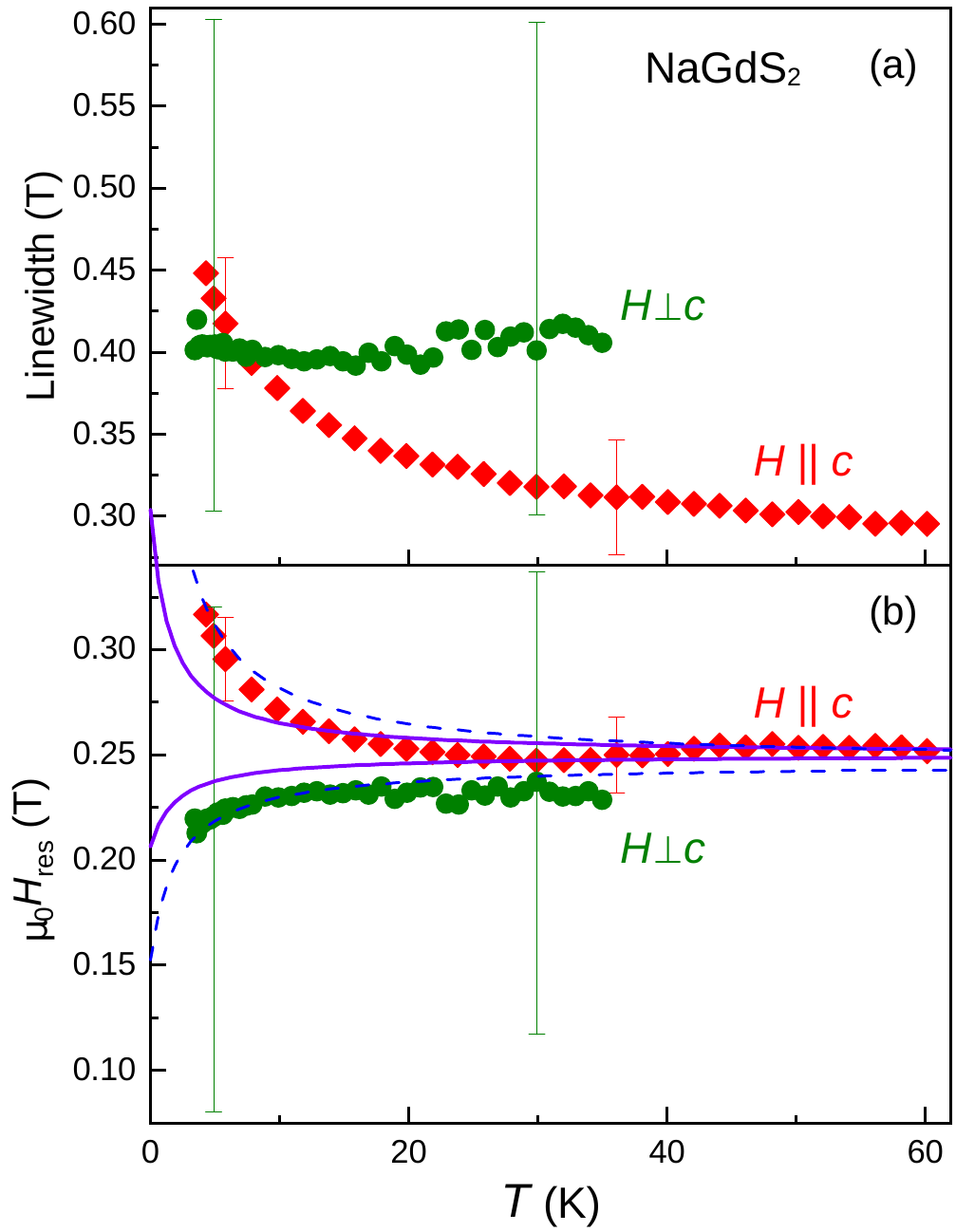}
\end{center}
\caption{Temperature dependence of (a) the ESR linewidth and (b) resonance field for $H\| c$ and $H\perp c$. At 60\,K, the resonance field corresponds to an effective $g$-factor $g_\|=2.6(7)$. Solid lines denote a molecular-field description of $\mu_0H_{\rm res}$, using Eqs. (\ref{eq:1}) and (\ref{eq:2}) with $\theta_\bot$ and $\theta_\|$ as determined by $\chi_{\rm DC}$ [see Fig. \ref{sus} (b)]. Dashed lines are results from the same equations based on simulated Curie-Weiss temperatures (see discussion).  
}
\label{Fig2ESR}
\end{figure}

Figure \ref{Fig2ESR}(a) shows the temperature dependence of the ESR linewidth $\Delta H$, while Fig. \ref{Fig2ESR} shows that of the resonance field $H_{\rm res}$ for $H\| c$ and ${H}\bot{c}$. 
The small signal amplitude relative to the background signal causes large error bars for $H\bot c$, see Fig. \ref{FigESR1}(a). 

%

Towards low temperatures, the growing influence of spin correlations lead to an increase of  $\Delta H$ as well as to increasing internal fields, which affect the resonance field $H_{\rm res}$. The latter qualitatively resembles the behavior of the Yb$^{3+}$ ESR in NaYbS$_{2}$ \cite{sichelschmidt19a}, for which a molecular-field model \cite{huber09a} of the anisotropic Yb--Yb interaction provided a reasonable description. 
Adapting this molecular-field model to the temperature dependence of the resonance field results in
\begin{eqnarray}
\label{eq:1}
H_{\bot\rm res}(T) &=& H_{\bot\rm res}^0 \left(1+\frac{\theta_\bot - \theta_\|}{T - \theta_\bot}\right)^{-\frac{1}{2}},\\
\label{eq:2}
H_{\|\rm res}(T) &=& H_{\|\rm res}^0  \left(1-\frac{\theta_\bot - \theta_\|}{T - \theta_\|}\right)^{-1}.
\end{eqnarray}
Here, $\theta_\bot$ and $\theta_\|$ are the Curie-Weiss temperatures as derived from our dc-susceptibility data (see Fig. \ref{sus}). Using $\mu_0H_{\bot\rm res}^0=\mu_0H_{\|\rm res}^0=0.25$\,T results in a reasonable description of the observed temperature dependence of the resonance field [solid lines in Fig. \ref{Fig2ESR}(b)]. 

Hence, the weak temperature dependence of $H_{\rm res}$ confirms the $\theta_\bot$ and $\theta_\|$ values, which we extracted from our dc-susceptibility data (Fig. \ref{sus}).
We are also able to explain our magnetization data at about 5\,K by a modified $g$-factor of $g_{\parallel}=2.2(0)$ under the assumption that internal fields are weak enough. The calculated values are $\mu_{\rm eff}$ = 8.7\,$\mathrm{\mu_B}$ and $M_{\rm sat}$ = 7.7\,$\mathrm{\mu_B}$ based on this $g$-factor using the equations mentioned in section A. These results are close to the values measured by magnetization. 
The ESR measurements thus confirm the anisotropy of NaGdS$_2$.

\subsection{$^{23}$Na nuclear magnetic resonance}
\begin{figure}[H]
    \centering
    \includegraphics[width=0.47\textwidth]{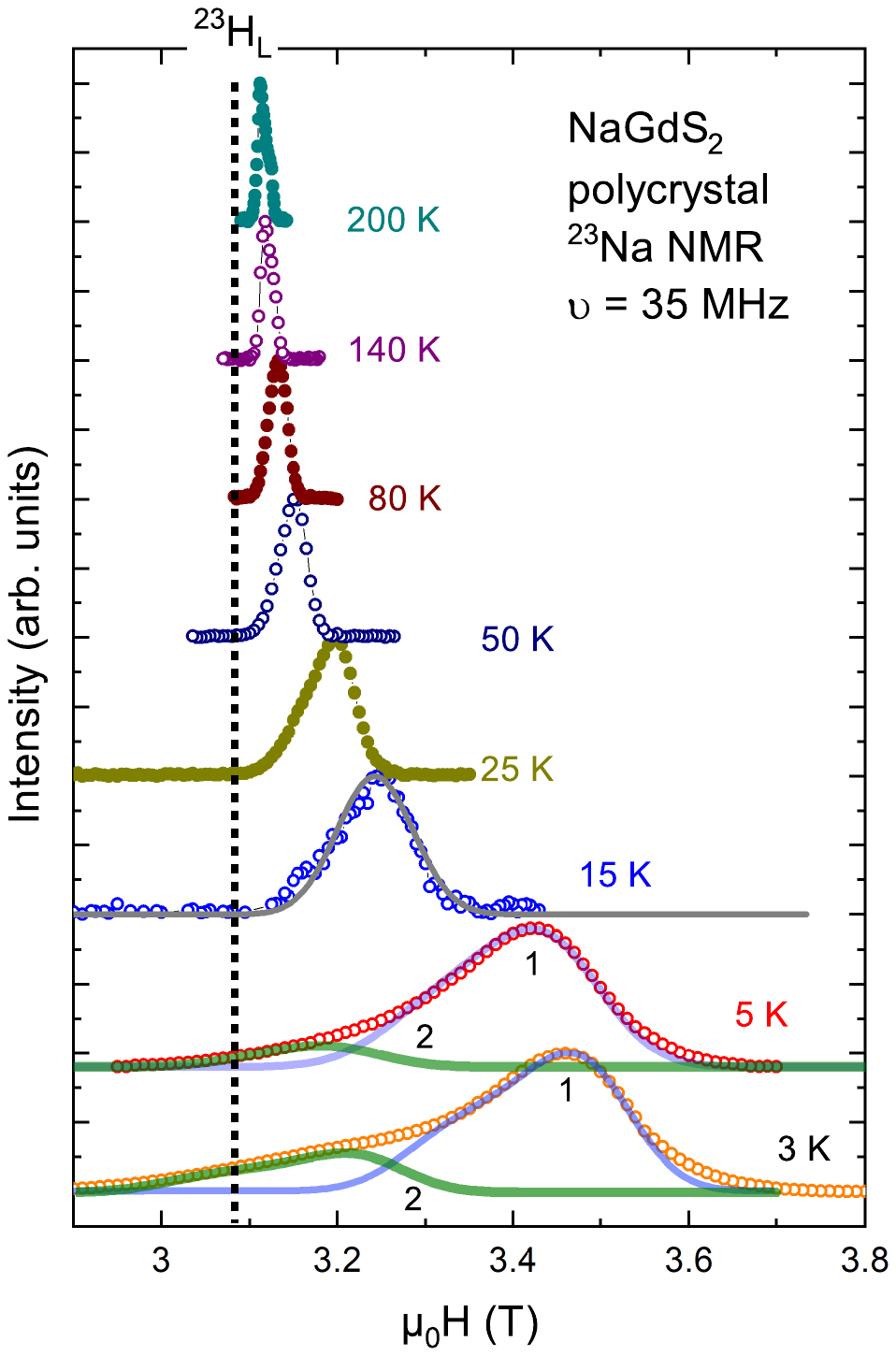}
    \caption{$^{23}$Na field-sweep NMR spectra taken at 35\,MHz.
The solid lines represent simulations (see text).  The vertical dotted line indicates the Larmor field for
$^{23}$Na.}
    \label{fig:NMR1}
\end{figure}

Figure \ref{fig:NMR1} shows
of field-sweep powder NMR spectra at selected temperatures. The powder spectra are described using an anisotropic shift tensor to determine the components for the
$a$, $b$, and $c$ directions. We determined the isotropic part of the Knight shift,
$^{23}K$ by $^{23}K_\mathrm{iso} = (K_a + K_b + K_c)/3$.

\begin{figure}[H]
    \centering
    \includegraphics[width=0.47\textwidth]{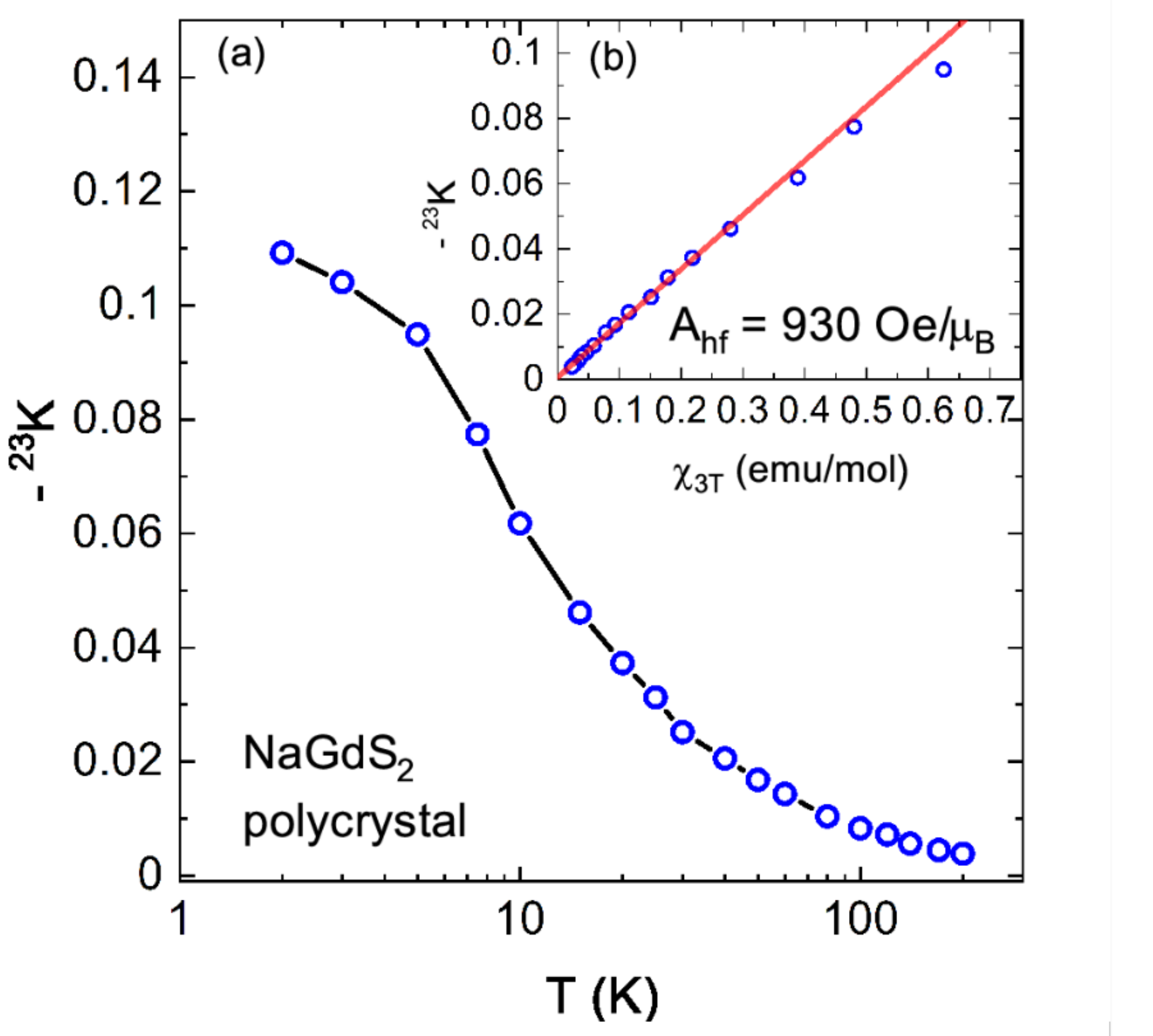}
    \caption{(a) Temperature dependence of the negative isotropic $^{23}$Na lineshift $^{23}K$ in NaGdS$_2$ estimated from the 35\,MHz field-sweep NMR data (Fig. \ref{fig:NMR1}). (b) Knight shift vs. susceptibility ($^{23}K$--$\chi$ plot). Note that here the susceptibility at 3\,T was used. The red line shows a fit using Eq. (4) with temperature as implicit parameter.}
    \label{fig:NMR2}
\end{figure}

 While we observed no large anisotropy at high temperatures, the lines below about 10\,K, described by two NMR lines, become very broad. One NMR line (labeled 1 in the following) shows an anisotropic broadening below 25\,K. Additionally a second NMR signal 2 emerges. The main NMR line 1 has a negative hyperfine field and a negative shift. The line 2 has a positive (ferromagnetic) component. We mentioned the evolution of magnetic fluctuations above in connection with the ESR results (Fig. \ref{FigESR1}). Overall, an explicit description of the NMR spectrum below 10\,K is not possible. Therefore, we only used data above 10\,K to analyze the hyperfine coupling.  The ``Clogston-Jaccarino'' plot ($^{23}K$--$\chi$ plot, Fig. \ref{fig:NMR2}) relates the NMR shift to the bulk susceptibility. 
The observed isotropic shift is related to the magnetic bulk susceptibility by
\begin{equation}
^{23}K_\mathrm{iso}(T) = K_0 + A_\mathrm{hf} \chi(T)/N_\mathrm{A}\mu_\mathrm{B},
\end{equation}
where $A_\mathrm{hf}$ is the hyperfine coupling constant, $\chi$ is the bulk
magnetic susceptibility (at the NMR field) and $K_0$ is a residual temperature-independent
contribution. Using this equation, we obtain the hyperfine coupling constant $A_\mathrm{hf} =
930$\,Oe/$\mu_B$ and a negligibly small $K_0$. 
The value of the hyperfine coupling constant in NaGdS$_{2}$ is roughly twice as large as in the NaYb$X_{2}$ compounds (above 100\,K, where the $S$ = 7/2 state dominates the magnetization). The NMR data on the polycrystal can be summarized as follows: The local NMR susceptibility (NMR shift) follows the bulk susceptibility down to approximately 10\,K. At lower temperatures, a magnetic anisotropy develops and a new NMR line appears. We interpret this as an indication for the evolution of short-range order in the system, possibly with ferromagnetic origin, typically associated to dipolar interactions. Results taken at lower fields (9.5\,MHz, not shown) are consistent with the 35\,MHz data. 

\subsection{AC susceptibility at mK temperatures}

\begin{figure}[H]
  \begin{center}
  \includegraphics[width=\columnwidth]{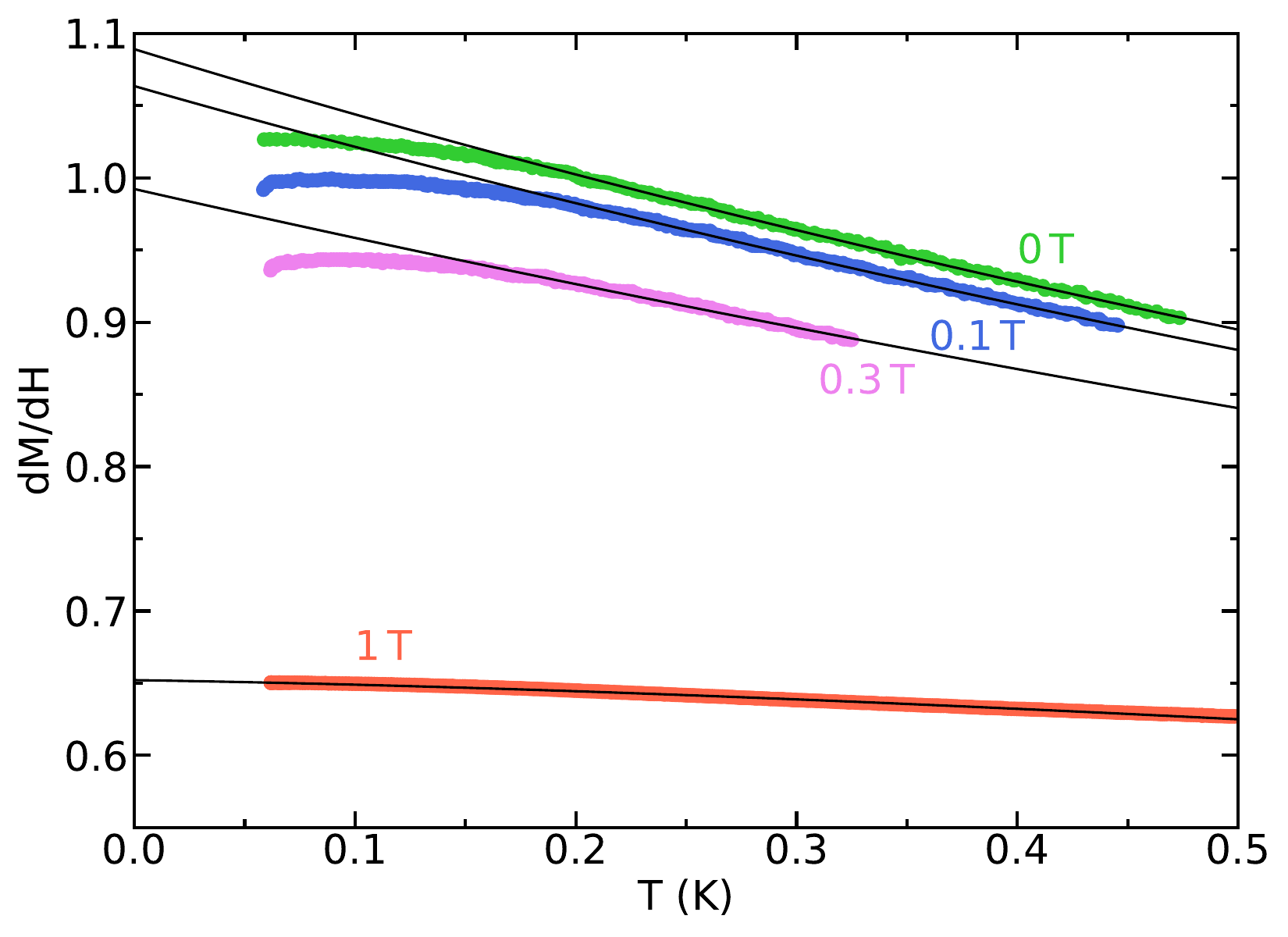}    
  \end{center}
  \caption{Temperature-dependent AC susceptibility measured using a $\mu T$ ac excitation and a frequency of 1123\,Hz in zero and various static magnetic fields. 
  The black lines indicate ideal paramagnetic behavior.}
  \label{acs}
\end{figure}

The ESR and NMR measurements above 2\,K evidenced short-range magnetic correlations towards low temperatures, so magnetic order might appear at sub-kelvin temperatures. 
For this reason, we performed ac-susceptibility measurements at this temperatures in the orientation $H\parallel c$,  perpendicular to the Gd$^{3+}$-layers. 
First, we performed measurements at various frequencies in zero static magnetic field (123\,Hz, 525\,Hz, and 1123\,Hz). A difference in the result could not be found, so we continued with experiments at the optimized frequency of 1123\,Hz (see Fig. \ref{acs}). We compared the result with the differential susceptibility $\chi_{\rm dif} = \frac{\mathrm{d}M}{\mathrm{d}H}$, which was calculated from paramagnetic Brillouin functions, since the Curie law does not apply completely at mK temperatures. 
We observe a deviation from the paramagnetic Curie law at about 180\,mK in zero field. Below this temperature, the susceptibility stays constant and is lower than the paramagnetic susceptibility, indicating dominant antiferromagnetic contributions. However, a sharp cusp is absent, so that a distinct transition representing magnetic order cannot be verified.

To track the deviation from paramagnetism, we performed measurements at different magnetic fields. The deviation from the fully paramagnetic curve remains at about 180\,mK for 0.1\,T and barely shifts down to about 160\,mK for 0.3\,T. This shift is again characteristic for an antiferromagnetically dominated behavior. At 1\,T the material is finally saturated, as suggested by the magnetization at 0.5\,K (see Fig. \ref{mag}).

\subsection{Thermal expansion and magnetostriction}
\begin{figure}[H]
  \begin{center}
  \includegraphics[width=\columnwidth]{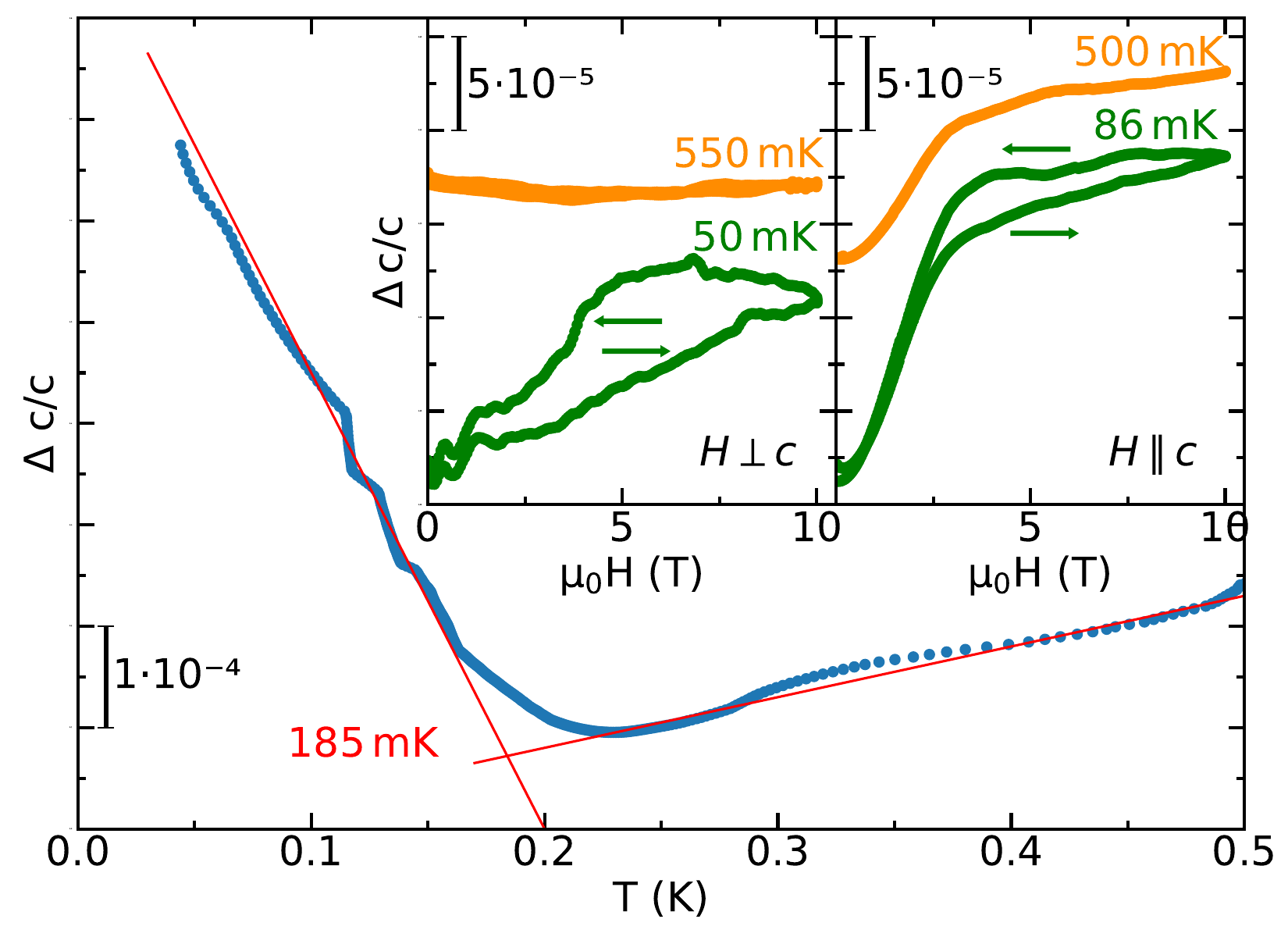}
  \end{center}
  \caption{Thermal expansion along the crystallographic $c$ direction in zero field. The insets show the magnetostriction above and below the slope change occurring at 185\,mK for both orientations. }
  \label{Ausdehnung}
\end{figure}

To further investigate possible magnetic aspects, we performed thermal expansion and magnetostriction measurements at very low temperatures. 
Due to the plate-like shape of the single crystals, we could only measure length changes along the $c$ direction ($\Delta c/c$). We applied magnetic fields parallel (longitudinal geometry) and perpendicular (transversal geometry) to $c$. We measured the relative length change at zero field during cool down and found reproducibly a broad change of slope at about 185$\,$mK, close to the temperature at which the deviation from paramagnetic susceptibility occurs
(Fig. \ref{Ausdehnung}). The slope  $\frac{d}{dT}(\Delta c/c)$ changes from a negative at low temperatures to a smaller positive increase above about 185 mK. We assume that exchange striction is the dominant mechanism \cite{Magnetostriktiontheorie}. 
As a result, the distances between the magnetic sites decrease towards lower temperatures. It follows that the $ab$ plane shrinks. Due to the assumed conservation of volume, the $c$ direction should expand towards lower temperatures, which is what we observe below 185\,mK. Above this temperature, we see the thermal expansion due to the anharmonic lattice vibration and no magnetic contribution, which indicates paramagnetism. 

We further measured the magnetostriction for the two field orientations (Fig. \ref{Ausdehnung} insets). For temperatures below the slope change, we found hysteretic behavior, which may be caused by ferromagnetic contributions. Comparing both orientations, the hysteresis loop of the in plane direction has a larger opening than the loop for out-of-plane field. This indicates a stronger in-plane ferromagnetic contribution than out of plane, which is consistent with magnetization and ESR results.
In the paramagnetic phase, apart from the absence of hysteresis, the transversal magnetostriction curve ($H\perp c$) is almost independent of external magnetic field, whereas the longitudinal one ($H\parallel c$) is similar to the magnetization with a saturation at about 3-4\,T. The effect of parastriction \cite{bernath, bienkowski} can cause this behavior. 

\section{Discussion}
The Gd$^{3+}$-based magnet NaGdS$_2$ shows paramagnetic behavior with only weak anisotropy at higher temperatures, due to suppressed crystal-field effects [Fig. \ref{sus}(b)]. However, the Brillouin functions at 2 and 10\,K (black lines in Fig. \ref{mag}) show deviations from the measured magnetization at small fields. These differences illustrate incipient correlation effects. 
At low temperatures we indeed found magnetic correlations in ESR, NMR, and magnetization measurements. Susceptibility and dilatometry measurements
indicate increasing correlations below
about 180 mK as well. Negative Curie-Weiss temperatures indicate predominant antiferromagnetic exchange. However, signs for some ferromagnetic interactions could also be deduced from NMR, magnetization, and magnetostriction measurements, especially for in-plane magnetic fields. 

Ferromagnetism caused by dipolar interactions is large enough to induce such an anisotropic effect. A consideration of nearest neighbors supports this assumption. Nearest in-plane Gd$^{3+}$ neighbors have a distance of 4.02\,\r{A}, whereas the distance to the nearest out-of-plane neighbor is 7.00\,\r{A}. 

Low Curie-Weiss temperatures, fast saturation of magnetization data, and strong magnetic-field dependence of the specific heat suggest the presence of only weak exchange interactions. This leads to similar strengths of exchange and dipolar interactions, which may cause additional frustration

A more detailed description of the energetic conditions in NaGdS$_2$ is beyond the score of this work. A similar analysis as done for KBaGd(BO$_3$)$_2$ \cite{xiang2023dipolar} seems reasonable. We used the \textit{McPhase} program package \cite{mcphasemanual}, which provides a Monte Carlo analysis of magnetic interactions using a molecular-field approach. The magnetic Hamiltonian can be written as
\begin{equation}
  \mathcal{H} = \mathcal{H}_{\rm ex} + \mathcal{H}_{\rm dip},
\end{equation}
with Heisenberg exchange ($\mathcal{H}_{\rm ex}$) and dipole term ($\mathcal{H}_{\rm dip}$). With the coupling strengths $J_{\rm H}$ and $J_{\rm D}$, we may write 
\begin{equation}
      \mathcal{H}_{\rm ex} = J_{\rm H}\sum_{\rm <ij>} \boldsymbol{S_{\rm i}}\cdot \boldsymbol{S_{\rm j}},
\end{equation}
\begin{equation}   
  \mathcal{H}_{\rm dip} = -J_{\rm D}\sum_{\rm <ij>\in d}[\boldsymbol{S_{\rm i}}\cdot \boldsymbol{S_{\rm j}} - 3(\boldsymbol{S_{\rm i}}\cdot \boldsymbol{e_{\rm ij}})(\boldsymbol{S_{\rm j}}\cdot \boldsymbol{e_{\rm ij}})]/r_{\rm ij}^3.
\end{equation}
In these expressions, the summation runs over all spins  $S_{\rm i}$, $S_{\rm j}$, while $e_{\rm ij}$ and $r_{\rm ij}$ denote the direction and distance between the Gd$^{3+}$ ions i or j, respectively. 

We first determined the magnetic structure taking into account only dipolar interactions with 
\begin{equation}
  J_{D} = \frac{\mu_0}{4\pi}(g\mu_{\rm B})^2
\end{equation}
using Eq. (7) and summing up all long-range dipole-dipole-interactions up to $d$ = 23\,\r{A}.
Equation (8) uses $\mu_0$ as the permeability of the vacuum, $\mu_{\rm B}$ the Bohr magneton and $g$ the Landé factor. 
The simulation converges to a ferromagnetic structure of the Gd$^{3+}$ moments in the basal plane, as expected from dipolar interaction, and an antiferromagnetic stacking in the c direction with 120$^\circ$ rotation between neighboring planes. The susceptibility for magnetic fields $H\parallel c$ and $H \perp c$ (field in $a$ direction) results in Curie-Weiss temperatures of $\theta_{\parallel}^{dip}$= -1.1\,K and $\theta_{\perp}^{dip}$ = 0.5\,K, respectively.
These values are larger than those determined experimentally ($\theta_{\parallel}^{exp}$= -2.2\,K and $\theta_{\perp}^{exp}$ = -1.5\,K).
This difference, presumably, is due to the exchange interactions.

In regard of the temperature differences $\Delta\theta_{\rm k}$ ($\Delta\theta_{\rm k} = \theta_{\rm k}^{\rm cal} - \theta_{\rm k}^{\rm exp}, {\rm k}=\perp \rm{or} \parallel$), $J_H$ could be calculated by
\begin{equation}
  J_{\rm H} = -\frac{3k_{\rm B}\Delta\theta}{zS(S + 1)},
\end{equation}
where $z$ = 6 is the number of nearest neighbors in the basal plane and $S$ = 7/2.
Already $J_H$ = 52\,mK is sufficient to explain the experimental Curie-Weiss temperatures, which is a remarkably low value. Using equation (5), we obtain $\theta_{\parallel}^{dip}$= -2.6\,K and $\theta_{\perp}^{dip}$ = -1.0\,K.

These results are in good correspondence to our dc-susceptibility measurements. In the experimental case, the anisotropy due to dipolar contributions is reduced by additional quantum-fluctuations, which are not included in this simple model, and slight misalignments in the sample preparation. We have also described the field dependence of the ESR resonance fields by the simulated Curie-Weiss temperatures. Using equations (2) and (3), we were able to obtain the dashed lines in Fig. \ref{Fig2ESR}, also resembling the measurement.

In that sense, the delafossite NaGdS$_2$ represents one of the rare compounds 
with competing exchange mechanisms (dipolar and Heisenberg) of almost the same energy scale ($J_H$ = 52\,mK and $J_D$/(4.02\,\r{A})$^3$ = 46\,mK). 
Furthermore, it is noteworthy that dipolar interactions lead to a clear anisotropy in a Gd$^{3+}$ compound without single-ion anisotropy, as also found in the ESR measurements. Both, theoretical and experimental Curie-Weiss temperatures used in equations (2) and (3) describe the ESR measurements well (see solid and dashed line in Fig. \ref{Fig2ESR}).

This magnetic behavior with strong dipolar interactions is not new for Gd-based triangular-lattice magnets \cite{xiang2023dipolar, GuoGd19, JescheGd22}. Even Yb-based triangular-lattice magnets can have appreciable dipolar interactions \cite{BagBYB}.  

About a possible order at lower temperatures, we only can speculate. Based on the calculations above, we would expect a stripe-antiferromagnetic order in plane, despite the isotropy of Gd$^{3+}$, caused by strong dipolar interactions.

\section{Summary}
NaGdS$_2$ shows paramagnetic behavior with weak anisotropy at high temperatures. However, at lower temperatures sizable magnetic correlations appear in ESR, NMR, and specific-heat data. In particular, we could find signs of antiferromagnetic Heisenberg-type and ferromagnetic dipolar contributions. Moreover, the results indicate that the dipolar interaction leads to an anisotropy at low temperatures. This is seen in stronger ferromagnetic effects for in-plane than out-of-plane magnetic fields. 

Susceptibility and dilatometry measurements at very low temperatures show increasing magnetic fluctuations below about 180\,mK. 
This is in accordance to only weak exchange, which corresponds to significantly lower Curie-Weiss temperatures compared to related delafossites. Also magnetization, magnetostriction, and specific-heat data support this fact. Calculations reveal that dipolar and exchange energies are of the same order of magnitude. This leads to the absence of magnetic order down to very low temperatures.

This combination of absent order, weak exchange interactions, and strong dipolar interactions is new for rare-earth delafossites and hints to spin-liquid character also for NaGdS$_2$.

\begin{acknowledgments}
We acknowledge support from Deutsche Forschungsgemeinschaft (DFG) through SFB~1143 (Project-id 247310070) and the W\"urzburg-Dresden Cluster of Excellence on Complexity and Topology in Quantum Matter--$ct.qmat$ (EXC 2147, Project No.\ 390858490). We as well acknowledge the support of the HLD at HZDR, member of European Magnetic Field Laboratory (EMFL).
\end{acknowledgments}

\section*{Data availability}

Data is available upon reasonable request from the contact authors; data underpinning this work are also available from Ref. \cite{Ng_TUD}.

\bibliography{Ng_cites}

\end{document}